# Laser-Assisted Chemical Modification of Monolayer Transition Metal Dichalcogenides

Tariq Afaneh [1], Prasana K. Sahoo [1], Igor A. P. Nobrega [1], Yan Xin [2] and Humberto R. Gutiérrez [1]*

[1] *Department of Physics, University of South Florida, Tampa, Florida 33620, USA.*

[2] *National High Magnetic Field Laboratory, Florida State University, Tallahassee, Florida 32310, USA.*

* Corresponding Author: humberto3@usf.edu

**Abstract**:

Laser-assisted chemical modification is demonstrated on ultra-thin transition-metal dichalcogenides (TMDs), locally replacing selenium by sulfur atoms. The photo-conversion process takes place in a controlled reactive gas environment and the heterogeneous reaction rates are monitored via *in situ* real-time Raman and photoluminescence spectroscopies. The spatially localized photo-conversion resulted in a heterogeneous TMD structure, with chemically distinct domains, where the initial high crystalline quality of the film is not affected during the process. This has been further confirmed via transmission electron microscopy as well as Raman and Photoluminescence spatial maps. Our study demonstrates the potential of laser-assisted chemical conversion for on-demand synthesis of heterogeneous two-dimensional materials with applications in nanodevices.

Laser-assisted modifications have emerged as an alternative and reliable technique for local tailoring of the chemical, structural, optical and electrical properties in a variety of nanomaterials.[1] In particular, for layered and two-dimensional materials, [2-6] laser-thinning of layered $MoS_2$ [2, 3], $WS_2$ [2] and $WSe_2$ [6] have proven to be an efficient way to produce on-demand monolayer films. Additionally, laser induced site-specific doping of ultrathin $MoS_2$ and $WSe_2$ in a phosphine environment [4] has been demonstrated, facilitating the localized modification of the intrinsic photoluminescence as well as the electrical response in a 2D device configuration. Enhancement of the electrical conductivity of $WSe_2$ monolayer was observed during the laser assisted oxygen-passivation of chalcogen vacancies.[6] Local oxidation using a focused laser beam was also demonstrated to be effective in modifying the photo response of phosphorene-based 2D devices [5]. Additionally, a laser patterning technique has been employed to fabricate micro-supercapacitors using paintable graphene [7] and $MoS_2$ film [8]. The success of the above-mentioned experiments suggests that post-growth laser-assisted methods can become a reliable route to tailor the physicochemical properties of 2D materials. However, other than inducing evaporation, oxidation or doping, the potential of post-growth laser assisted method is yet to be verified for in-situ changing of the chemical composition of transition metal dichalcogenides, such as creating localized ternary alloys, or even completely replacing the chalcogen atoms. Here, we report the successful laser-induced chemical modification of suspended TMD monolayer films via local exchange of the chalcogen atoms; selenides to sulfides. With the proposed method, total or partial replacement of the chalcogen atoms was achieved, in both $WSe_2$ and $MoSe_2$ suspended films. The time constants associated with the different photo-chemical mechanisms involved in the conversion process were studied by *in situ* monitoring the Raman and photoluminescence spectra of the samples. Our results suggest that post-growth laser-induced chemical modifications could be considered as an alternative route for the fabrication of spatially localized ternary alloys and in-plane 2D heterostructures in a controlled gas environment.

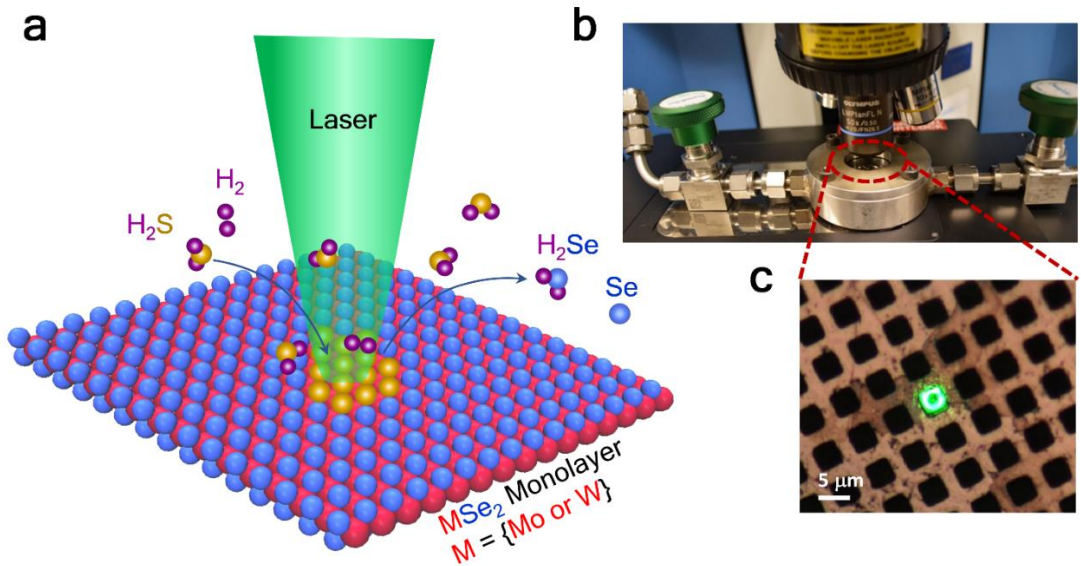

**Figure 1.** (**a**) Schematics of the laser-induced chalcogen atom exchange under an $H_2S$ reactive atmosphere. (**b**) Image of the experimental setup showing the mini-chamber under the microscope-based Raman spectrometer. (**c**) Optical image of the laser beam (green) focused on the TMD film suspended on a TEM grid.

A schematic representation of the photo-induced chemical reaction as well as the experimental setup are shown in Figure 1. The process takes place within a home-made sealed stainless steel mini-chamber with a controlled gas environment, $H_2S$ in this case. An optically transparent quartz window allows to focus the laser beam on the sample surface ($WSe_2$ or $MoSe_2$) inside the chamber. The laser irradiation plays a double role in creating selenium vacancies via local heating and dissociating the $H_2S$ gas molecules which further take part in the chemical exchange (Figure 1a). The wavelength of the laser, power, exposure time, and the composition of the reactive atmosphere are the main parameters that directly affect the photo-conversion process. We evaluated the effect of different laser wavelengths (632 nm, 532 nm and UV light) on the photo-conversion yield. The excitation wavelength of 532nm produced the best results, whereas no chemical conversion was observed with the 632 nm laser. Exposure to UV light, on the other hand, induced uncontrolled and delocalized dissociation of the $H_2S$ molecule resulting in sulfur deposition over the entire sample surface and on the quartz optical window. Hydrogen and sulfur generation by photolysis of H$_2$S gas molecules under UV radiation have been extensively studied in the past. [9]

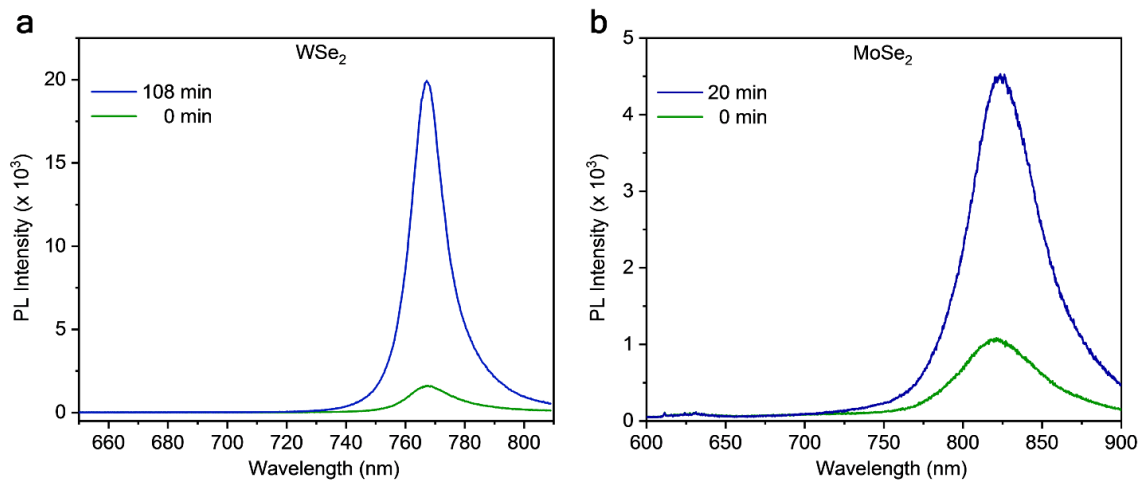

**Figure 2.** PL spectra of as-grown (green curves) and laser-healed (blue curves) samples of (**a**) WSe$_2$ and (**b**) MoSe$_2$. All the spectra were taken on suspended films.

The laser power was found to be a critical parameter. Very high laser power (> 1 mW) irreversibly damage the sample crystal structure. On the other hand, low laser power (< 0.3 mW) do not contribute significantly to the removal of selenium atoms and subsequent creation of chalcogen vacancy that are needed for the chemical exchange. Nevertheless, samples irradiated with low laser powers (~0.3 mW) improved the photoluminescence efficiency without notable shift of the PL peak position (Figure 2), suggesting that although no significant chemical conversion happened, the number of non-radiative recombination centers (e.g. vacancies and defects) was reduced or passivated for this range of laser power. Both WSe$_2$ (Figure 2a) and MoSe$_2$ (Figure 2b) monolayers displayed increased PL intensity after several minutes of laser exposure in the presence of H$_2$S atmosphere. Atomic healing of chalcogen vacancies at low laser powers, in the presence of oxygen, have been previously reported [6]. In our case, the H$_2$S rich environment facilitates the incorporation of sulfur atoms to passivate the selenium vacancies already existing in the as-grown TMD material. Small amounts of sulfur (< 1%) are not expected to significantly affect the PL peak position; however, in some samples, a small PL peak shift (~50 meV) to higher energies was observed. Although this small shift could be associated with a transition from trions to free excitons due to neutralization of defect-related doping, alloying with low sulfur composition cannot be ruled out. $H_2S$ dissociates at high temperatures (>900 °C);[10] hence the increase in sample temperature due to irradiation with low laser powers might not be sufficient to dissociate the $H_2S$ molecule. However, the selenium vacancies already present in the CVD-grown film can act as localized centers to catalyze the $H_2S$ molecule dissociation. Previous reports have also shown the efficient dissociation of $H_2S$ molecule at significantly lower temperatures in the presence of catalysts. [11]

An intermediate laser power of 0.7 mW was found to be optimal for inducing photo-chemical conversion without notable damage to the crystalline structure. Using this laser power, two different experiments were conducted with distinct reactive atmosphere, the first experiment with only $H_2S$ gas and the second with a mixture of $H_2S$: $H_2$+Ar (4:1 vol.). For fixed conditions of laser power, wavelength and TMD film thickness, it is reasonable to assume that the Raman peak intensity of the different phonon modes is directly proportional to the number of metal-chalcogen chemical bonds. This allows *in situ* monitoring of the temporal variation of each chalcogen atom content through the time-dependent Raman intensity of its corresponding phonon modes. In multilayered $WSe_2$, the $A_{1g}$ and $E^1_{2g}$ phonon modes exhibit a small frequency difference (<3 cm$^{-1}$), for a monolayer these two modes overlap and become degenerated at 250 cm$^{-1}$,[12] for clarity in this manuscript we will refer to this peak simply as $A_{1g}$. In $WS_2$ the frequency of the $A_{1g}$ phonon mode is around 420 cm$^{-1}$ and is sensitive to the number of layers. [12, 13] For alloys ($WSe_{2(1-x)}S_{2x}$) this frequency can vary between 402 cm$^{-1}$ and 420 cm$^{-1}$ as a function of x.[14] The $E^1_{2g}$ (~355 cm$^{-1}$) and *2LA* (~350 cm$^{-1}$) modes in $WS_2$ are very close in frequency and their peaks often overlap, their positions are not significantly affected by number of layers[13] or composition,[14] but the intensity of the *2LA* mode is higher when the excitation wavelength is close to a double resonant condition.[13]

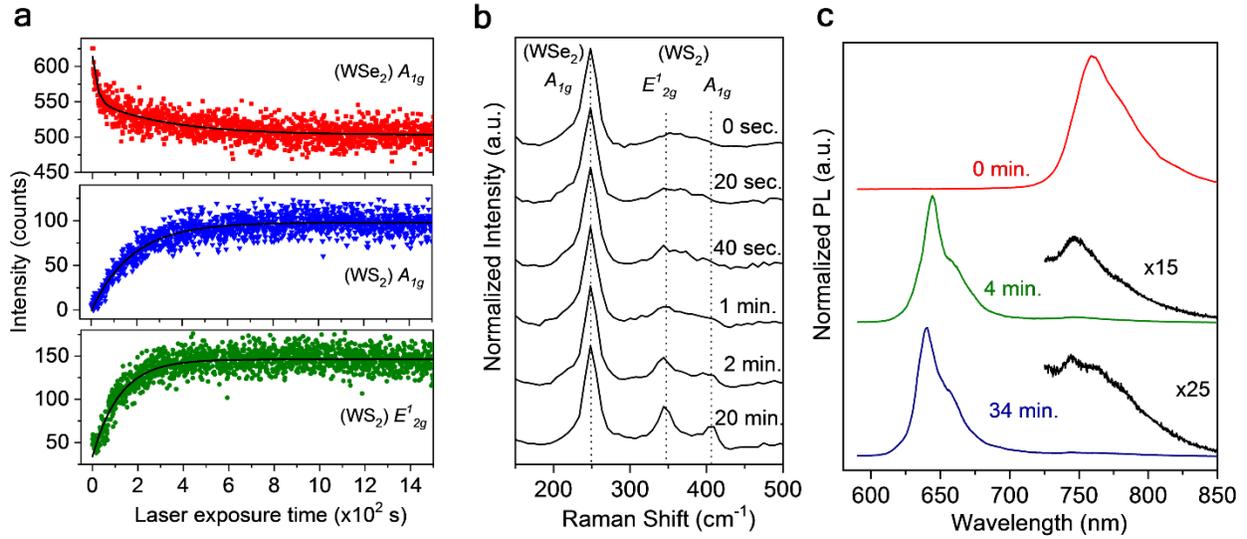

**Figure 3**. (**a**) Time-dependent evolution of the Raman intensity peaks for the WSe$_2$ (A$_{1g}$) and WS$_2$ (A$_{1g}$ and E$^1_{2g}$) under laser irradiation (532 nm, 0.7 mW) in a H$_2$S atmosphere. (**b**) Raman spectra for different exposure times. (**c**) Time evolution of the PL spectra for three different times, the black curves are a magnified section of the underlying spectrum to better visualize the low intensity bands between 730 – 800 nm.

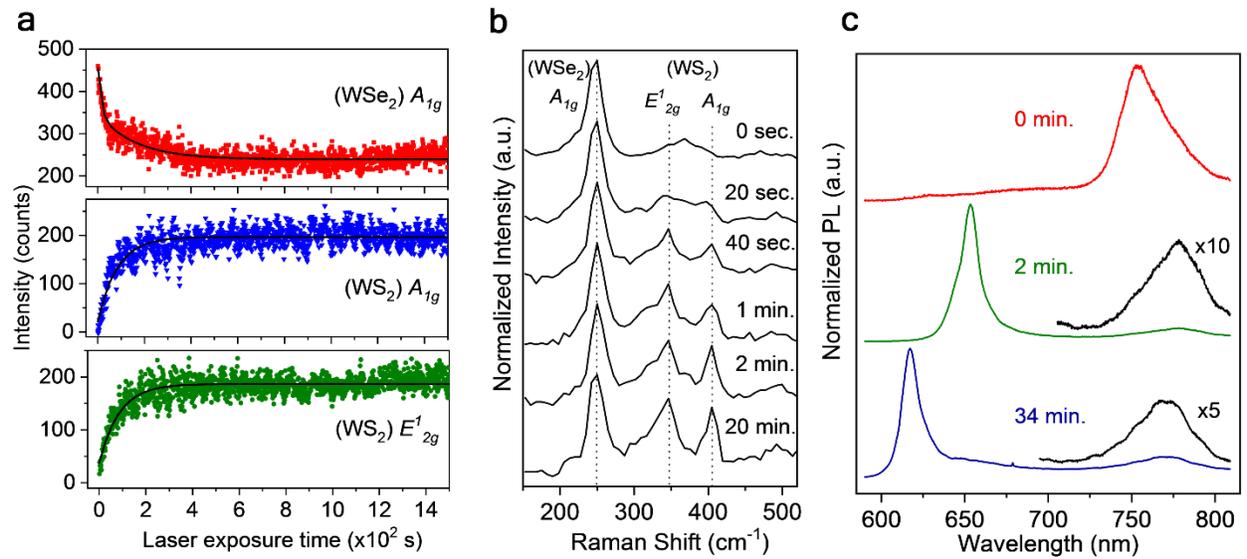

**Figure 4**. (**a**) Time-dependent evolution of the Raman intensity peaks for the WSe$_2$ (A$_{1g}$) and WS$_2$ (A$_{1g}$ and E$^1_{2g}$) under laser irradiation (532 nm, 0.7 mW) in a H$_2$S+H$_2$+Ar atmosphere. (**b**) Raman spectra for different exposure times. (**c**) Time evolution of the PL spectra for three different times, the black curves are a magnified section of the underlying spectrum to better visualize the low intensity bands between 730 – 800 nm.

Figures 3a and 4a show the time evolution of the Raman peaks intensities using $H_2S$ and $H_2S:H_2$+Ar environments, respectively. In both experiments, the intensity of the $A_{1g}$ phonon mode of $WSe_2$ exponentially decreases over time suggesting a reduction in the number of W-Se chemical bonds and hence the creation of Se vacancies. Contrary, the $A_{1g}$ and $E^1_{2g}$ phonon modes corresponding to $WS_2$, increased with time, which can be associated with subsequent incorporation of sulfur atoms into the TMD crystal lattice. As mentioned above, slight variations of the $A_{1g}$ peak position with respect to previous reports[13] can be attributed to alloying effects[14] during the photo-conversion process. The time-dependent data for the $WSe_2$ and $WS_2$ Raman intensities were best fitted to a double exponential decay (equation 1) and a single exponential function (equation 2), respectively.

$$I_{WSe_2} = (I_0 - I_{min})\left( e^{-\frac{t}{\tau_1}} + e^{-\frac{t}{\tau_2}} \right) + I_{min} \qquad (1)$$

$$I_{WS_2} = I_{max}\left( 1 - e^{-\frac{t}{\tau_3}} \right) \qquad (2)$$

An exponential time-dependence of the reactant concentration *[C]* is a fingerprint for first-order chemical reactions, where the rate of change of the concentration (*d[C]/dt*) is directly proportional to *[C]*, we refer the reader to the supporting information for a detailed solution of the differential equations. The double exponential decay in the $WSe_2$ Raman peak intensity, with different time constants ($\tau_1$ and $\tau_2$), suggests that two competing mechanisms for the creation of Se-vacancies can occur simultaneously: *i)* direct thermal removal (or evaporation) of Se atoms creating gaseous selenium, and *ii)* thermally-activated chemical reaction between selenium atoms in the TMD and the hydrogen molecules in the gas environment, releasing $H_2Se$ molecules. Since the shortest time constant ($\tau_1$) does not change significantly with the gas environment (see table 1), $\tau_1$ could be related to the first mechanism (thermal evaporation of Se). The second time constant *($\tau_2$)* is reduced to half when hydrogen is added to the gas environment, suggesting that $\tau_2$ characterizes the second mechanism. Even when only H₂S is used as reactive atmosphere, the hydrogen released after dissociation of H₂S molecules will be available to react with Se atoms creating new vacancies, but in less proportion. In fact, the presence of hydrogen accelerates the formation of Se vacancies, as indicated by the reduction of the time constant ($\tau_3$) associated with the sulfur incorporation (see table 1). Notice that, as expected, the time constants obtained for both phonon modes ($A_{1g}$ and $E^1_{2g}$) of WS₂ have very close values. In both experiments, the intensity of the Raman peak for WSe₂ is never completely reduced to zero, one possible explanation for these results could be due to the Gaussian intensity distribution of the laser spot; outer regions of the spot might not receive sufficient power to completely activate the photo-conversion process, but still contribute to the collected Raman signal. Another reason could be the presence of few-layer WSe₂ fragments that are more stable and difficult to convert into WS₂ than the monolayer WSe₂.

**Table 1**. Time constants obtained from fitting the time-dependent data in Figure 3a and 4a, to eq.(1) and eq.(2).

| Gas | Phonon Mode | $\tau_1$ (s) | $\tau_2$ (s) | $\tau_3$ (s) |
|---|---|---|---|---|
| H₂S | (WSe₂) $A_{1g}$ | 19 ± 3 | 321 ± 27 | - |
|  | (WS₂) $A_{1g}$ | - | - | 179 ± 4 |
|  | (WS₂) $E^1_{2g}$ | - | - | 120 ± 3 |
| H₂S+H₂ | (WSe₂) $A_{1g}$ | 13 ± 2 | 162 ± 10 |  |
|  | (WS₂) $A_{1g}$ | - | - | 83 ± 3 |
|  | (WS₂) $E^1_{2g}$ | - | - | 86 ± 3 |

The PL spectra were acquired only at certain times during the experiments. The PL spectra of the suspended WSe₂ films used in the photo-conversion experiments (top of Figure 3c and 4c), shows a strong peak around 760 nm consistent with monolayer $WSe_2$, the asymmetric shape of the peak toward lower energies suggests that the film also contains a small fraction of bilayer regions. When the photo-conversion took place in pure $H_2S$ gas (Figure 3c), the highest PL peak shifted from 760 nm (1.63 eV) to 645 nm (1.92 eV) in 4 minutes, and after 34 minutes slightly shifted to 640 nm (1.94 eV). This suggests that the main chemical conversion happens during the initial 4 - 6 minutes of the process, which agrees with the time-dependent Raman data. The final position of the PL peak at 1.94 eV is consistent with a high sulfur content (x~ 0.9) $WSe_{2(1-x)}S_{2x}$ ternary alloy.[14] When $H_2$ was also present in the gas environment (Figure 4c), the PL peak position shifted to 653nm (1.90 eV) after 2 min., and at the end of the process (after 34 min) the PL peak position was at 617 nm (2.01 eV), suggesting a complete conversion from $WSe_2$ to $WS_2$ at the center of the laser spot. A weak PL band (magnified black curves in Figure 3c and 4c) was observed between 730 – 800 nm (1.70 – 1.55 eV) after the conversion experiments. This low energy band could be related to a combination of PL from small regions of bilayer $WSe_2$ and/or from alloyed regions, with low-sulfur content, in the outer ring of the laser spot that were exposed to lower power due to the Gaussian intensity profile of the laser, this is in agreement with the Raman observations discussed above.

The spatial PL intensity maps (Figure 5) confirm that the exchange of chalcogen atom only took place in a region confined to the laser spot. Figure 5b show a map of the intensity ratio for the main PL peaks of $WS_2$ and $WSe_2$ ($I_{2.00eV}$ / $I_{1.63eV}$), corresponding to the experiment performed in $H_2S+H_2+Ar$ atmosphere. In this map, regions where the PL emission from monolayer $WS_2$ is very intense appear in red, while the blue regions consist of unaltered $WSe_2$. Figure 5c show normalized PL spectra taken at different positions along the dashed line in Figure 5b. The position of the PL peak (615 nm) within the chemically converted region is consistent with pure $WS_2$ (~2.0 eV). Figures 5d, 5e and 5f show a similar analysis for the sample processed in the $H_2S$ atmosphere. In this case, the PL peak position within the photo-converted region varies between 641 – 652 nm (1.94-1.92 eV) which suggests the formation of a sulfur rich ternary alloy (0.8<x<0.9 ).[14] In general, the PL signal at the WS$_2$ converted-regions is stronger than the PL from the WSe$_2$ film, this is also consistent with the higher signal-to-noise ratio observed for the normalized PL spectra corresponding to WS$_2$ region (Figure 5c). The fact that the WS$_2$ obtained from the photo-conversion process is a strong PL emitter suggests that little or no damage was introduced in the crystal lattice; since high concentration of crystalline defects could had reduced the overall PL yield.

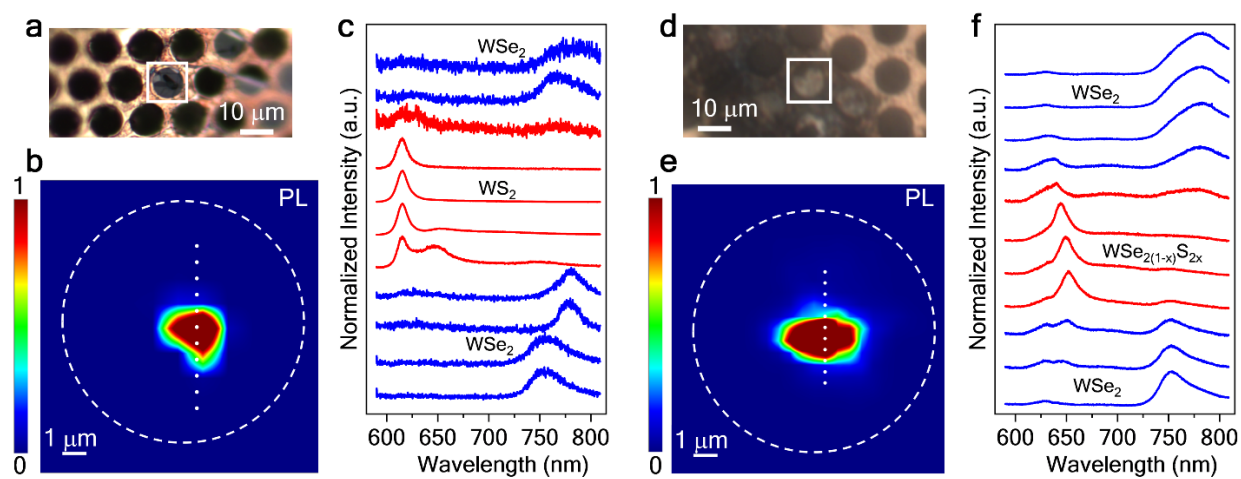

**Figure 5**. Spatially resolved PL data for the photo-converted samples in H$_2$S+H$_2$+Ar (**a**, **b** and **c**) and H$_2$S (**d**, **e** and **f**). (**a**) and (**d**) are optical image of WSe$_2$ monolayer films suspended on Ni TEM grids, the white squares indicates the grid holes where the photo-conversion experiments were performed. (**b**) and (**e**) are PL intensity maps of the ratio $PL_{WS2}/PL_{WSe2}$ and $PL_{WSe2(1-x)S2x}/PL_{WSe2}$, respectively, after conversion. The dashed white circles indicate the approximate position of the hole edge in the TEM grid. (**c**) and (**f**) are normalized PL spectra for different points along the straight dashed lines in (**b**) and (**e**), respectively.

The crystallinity and chemical composition, at the regions exposed to the laser, were further characterized by Z-contrast scanning transmission electron microscopy (STEM) (Figure 6a and 6b) using a high-angle annular dark field (HAADF) detector and electron energy loss spectroscopy (EELS) (Figure 6e and 6f). In Z-contrast STEM images with atomic-resolution, the relative scattered electron intensities at the different atomic positions can be used to identify the metal sites (W in this case) as well as different combinations of the chalcogen atoms (i.e. S$_2$, Se$_2$ or SSe) in monolayer TMDs. [15] [16] [17] Figure 6a shows an atomic-resolution STEM image from a region close to the center of the laser spot, revealing the high-quality single crystal structure that was preserved after photo-conversion in the $H_2S+H_2+Ar$ atmosphere. The corresponding electron intensity histogram (Figure 6c) displays only two intensity maxima associated to the $S_2$ sites (500-1000 a.u.) and W sites (2250-3000 a.u.), suggesting an almost complete replacement of selenium atoms by sulfur at the center of the laser spot. A region located around 500 nm away, also presents high crystalline quality (Figure 6b). Its electron intensity histogram (Figure 6d), however, contains four different features, three associated with the distinct chalcogen atoms site configurations ($S_2$, SeS and $Se_2$) and one feature for the W sites, characteristic of a $WS_{2x}Se_{2(1-x)}$ ternary alloy. As discussed above, the formation of a ternary alloy originates from partial conversion in the outer region of the Gaussian spot with less laser power.

    EELS spectra for the sulfur L-edge (Figure 6f), selenium L-edge and tungsten M-edge (Figure 6e) were also acquired at different positions relative to the laser spot; i.e. close to the center (green), at the outer ring (red) and in a region not exposed to the laser (black). We can see that as we approach the center, the EELS intensity of the sulfur L-edge increases (Figure 6f) while the selenium L-edge decreases (Figure 6e). The plural inelastic scattering background has been subtracted from these EELS spectra and the intensity of the sulfur and selenium signals are measured relative to the minimum intensity at the left of the spectra. The signal for the tungsten M-edge, however, overlaps with the tail of the selenium L-edge. Hence, the W signal is related to the height of the intensity step at the W M-edge, which is similar for the three regions. This is expected since, during the photo-conversion process, only the chalcogen atoms were replaced. Similar experiments of laser-assisted chemical modification were successfully conducted in suspended monolayers of $MoSe_2$, preliminary results are shown in the supplementary information. The conditions for $MoSe_2$ require further optimization, however, the fact that partial

or total photo-conversion was achieved in both materials (WSe$_2$ and MoSe$_2$) suggest that this approach could be implemented in a larger family of 2D materials.

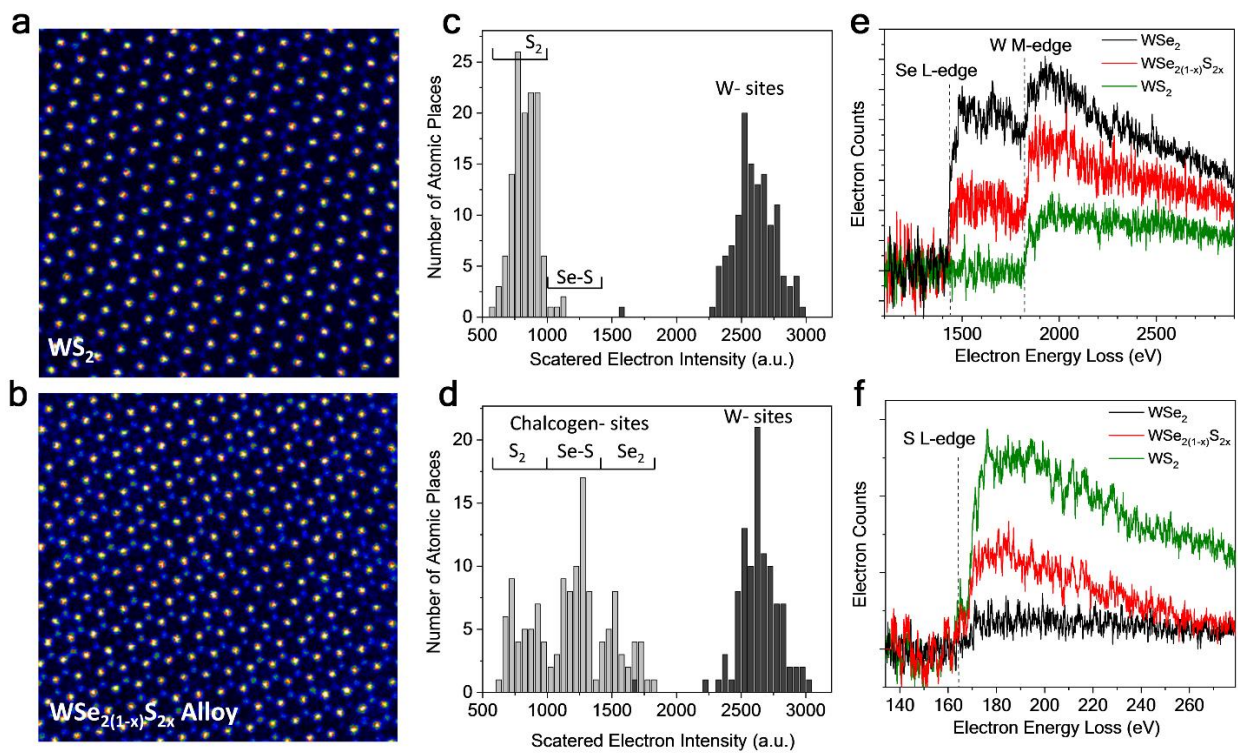

**Figure 6**. (**a**) and (**b**) False color atomic-resolution HAADF-STEM images at different positions in the chemically converted region, with complete chalcogen substitution (**a**) and partial chalcogen substitution (**b**). (**c**) and (**d**) are the corresponding electron intensity histograms of atom sites in (**a**) and (**b**), respectively. EELS spectra for (**c**) selenium L-edge and tungsten M-edge, and (**f**) sulfur L-edge at different positions along the chemically converted region exposed to the laser spot, center (green), edge (red) and out non-exposed (black).

In summary, a laser-assisted chemical conversion method for localized replacement of chalcogen atoms was demonstrated in suspended TMDs monolayer. The *in-situ* Raman characterization revealed that the photo-conversion process is dominated by first-order chemical reactions, and two different mechanisms for the formation of selenium vacancies were identified based on the time constants. The optimization of laser power, exposure time and reactive atmosphere is crucial to control the local composition and to preserve the crystalline quality of the films. Future research directions should combine laser-induced chemical conversion with lithographically defined patterns or shadow mask in order to control the size and shape of the modified domains. The method proposed in this work, is a new post-growth approach for creating ternary alloys as well as in-plane heterostructures of 2D chalcogenides with great potential for ultra-thin optoelectronic devices.

**Experimental section**

***TMDs Samples Preparation:*** Samples were prepared by Chemical Vapor Deposition (CVD). Either $WSe_2$ (99.8%) or $MoSe_2$ (99.9%) powders were used as solid precursors for the evaporation of the metal and chalcogen atoms. The synthesis was performed within a 1" quartz tube reactor placed inside a two-zone furnace in which the temperature of individual zones was controlled independently. A high purity alumina boat containing 100 mg of the solid precursors was placed at the hot zone of the furnace (~1060 °C), whereas $SiO_2$/Si substrates ($SiO_2$ thickness ~ 300 nm) were placed downstream at $T_{growth}$ ~750 - 800 °C for the growth of TMD films. Before and during the first 5 min. of the growth $N_2$ gas (200 sccm) was passed through a water bubbler in order to introduce small amount of water vapor in the system, after 5 min. the gas was switched to 200 sccm of $Ar+H_2$ (with 5% of $H_2$), this carrier gas was kept during the remaining 10 min. of growth and during the final cooling down. The presence of water vapor favors the formation of highly volatile species and the formation of high quality 2D TMD films.[18]

***Preparation of suspended TMD monolayer films:*** The TMD films were transferred from $SiO_2$/Si substrates onto Ni TEM grids using a well-established wet transfer process. [18, 19] In brief, the TMD films grown on Si/SiO2 substrates were spin-coated with PMMA (5 % vol) for mechanical support. Subsequently, the PMMA coated samples were immersed in a KOH solution (0.37g KOH / 10mL water, 0.65 M) to lift-off the PMMA+TMD film from the substrate. The typical time required for lift-off is around one hour. After this time, the PMMA+TMD transparent film, floating on the solution, is fished out with a glass slide and cut into smaller pieces (< 3mm diameter) that fit the TEM grids size where they were finally transferred. The PMMA support was then removed with acetone.

***Laser-assisted chemical conversion:*** The TMD ($WSe_2$ or $MoSe_2$) films suspended on the TEM grid where placed within a home-made stainless steel (SS 316) mini-chamber with an optically transparent quartz window. The chamber was evacuated with a vacuum pump and then filled with the desired reactive gas ($H_2S$ or $H_2S$: $Ar+H_2$) used as the source of S atoms during the photo-induced chemical exchange of chalcogen atoms (gauge pressure 10 psi). Finally, the small chamber was placed under the microscope of a Horiba LabRAM HR Evolution Raman system for simultaneous photo-conversion and *in situ* Raman spectrum acquisition. A 532 nm diode-laser was used in this study. An Olympus LMPLFLN 50x long working distance (LWD) objective lens with 10.6 mm working distance and Numerical Aperture (NA) of 0.50, as well as a 150 l/m grating were used in the experiments. During the defect-healing experiments, the laser power was 0.3 mW. However, for the photo-conversion experiments the laser power was increased to 0.7 mW. *Ex situ* Raman or PL mapping characterizations were performed with a 100x (NA = 0.9) objective lens, while keeping the laser power under 0.08 mW. For the experiments with excitation wavelength in the UV range, a UV lamp PSD Series Digital UV Ozone System, $\lambda$=185 nm and 254 nm, was utilized as the light source.

***Transmission Electron Microscopy:*** Scanning transmission electron microscopy (STEM) imaging was carried out in an aberration-corrected JEOL JEM-ARM200cF with a cold-field emission gun at 200kV. The STEM resolution of the microscope is 0.78 Å. The STEM images were collected with the JEOL High-Angle Annular Dark-Field (HAADF) detector using the following experimental conditions: probe size 7c, CL aperture 30μm, scan speed 32 μs/pixel, and camera length 8 cm, which corresponds to a probe convergence angle of 21 mrad and inner collection angle of 76 mrad.

## Acknowledgements

This work was supported by the National Science Foundation Grant DMR-1557434 (CAREER: Two-Dimensional Heterostructures Based on Transition Metal Dichalcogenides). TEM work was performed at the National High Magnetic Field Laboratory, which is supported by National Science Foundation Cooperative Agreement No. DMR-1644779, and the State of Florida.

# *Supporting Information*

## Laser-Assisted Chemical Modification of Monolayer Transition Metal Dichalcogenides

Tariq Afaneh [1], Prasana K. Sahoo [1], Igor A. P. Nobrega [1], Yan Xin [2] and Humberto R. Gutierrez [1†]

[1] *Department of Physics, University of South Florida, Tampa, Florida 33620, USA.*

[2] *National High Magnetic Field Laboratory, Florida State University, Tallahassee, Florida 32310, USA.*

### *Laser-induced chemical conversion from MoSe$_2$ to MoS$_2$:*

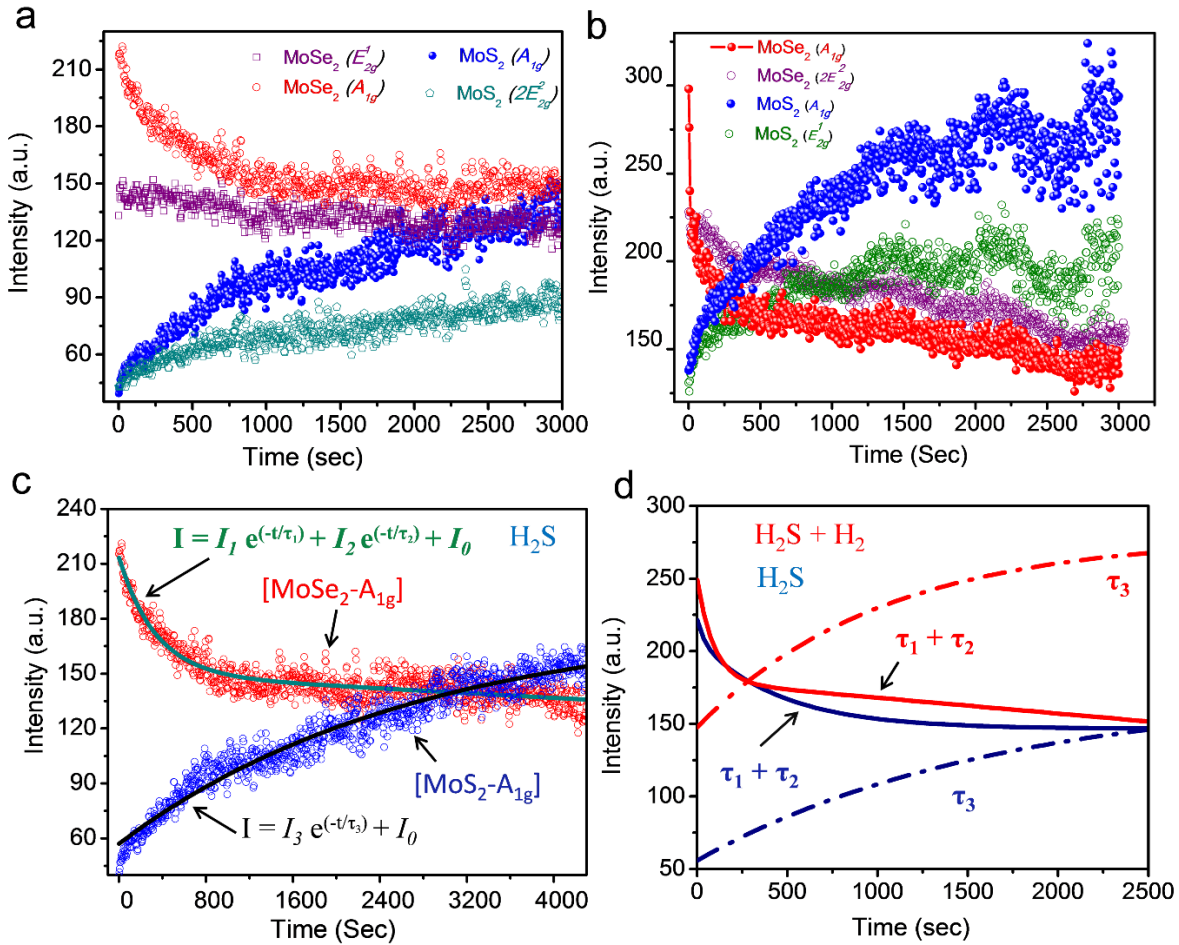

**Figure *S1*** Time-dependent evolution of the Raman intensity peaks for the MoSe$_2$ ($A_{1g}$ and $E^1_{2g}$) and MoS$_2$ ($A_{1g}$ and $E^1_{2g}$) under laser irradiation (532 nm, 0.7 mW); (a) in a H$_2$S atmosphere and (b) in a H$_2$S+H$_2$ atmosphere. (c) The fitting of MoSe$_2$ $A_{1g}$ and MoS$_2$ $A_{1g}$ peaks of figure (a). (d) The fitting of MoSe$_2$ $A_{1g}$ and MoS$_2$ A$_{1g}$ peaks under laser irradiation (532 nm, 0.7 mW) in a H$_2$S and H$_2$S + H$_2$ atmospheres (Supporting Information table S1)

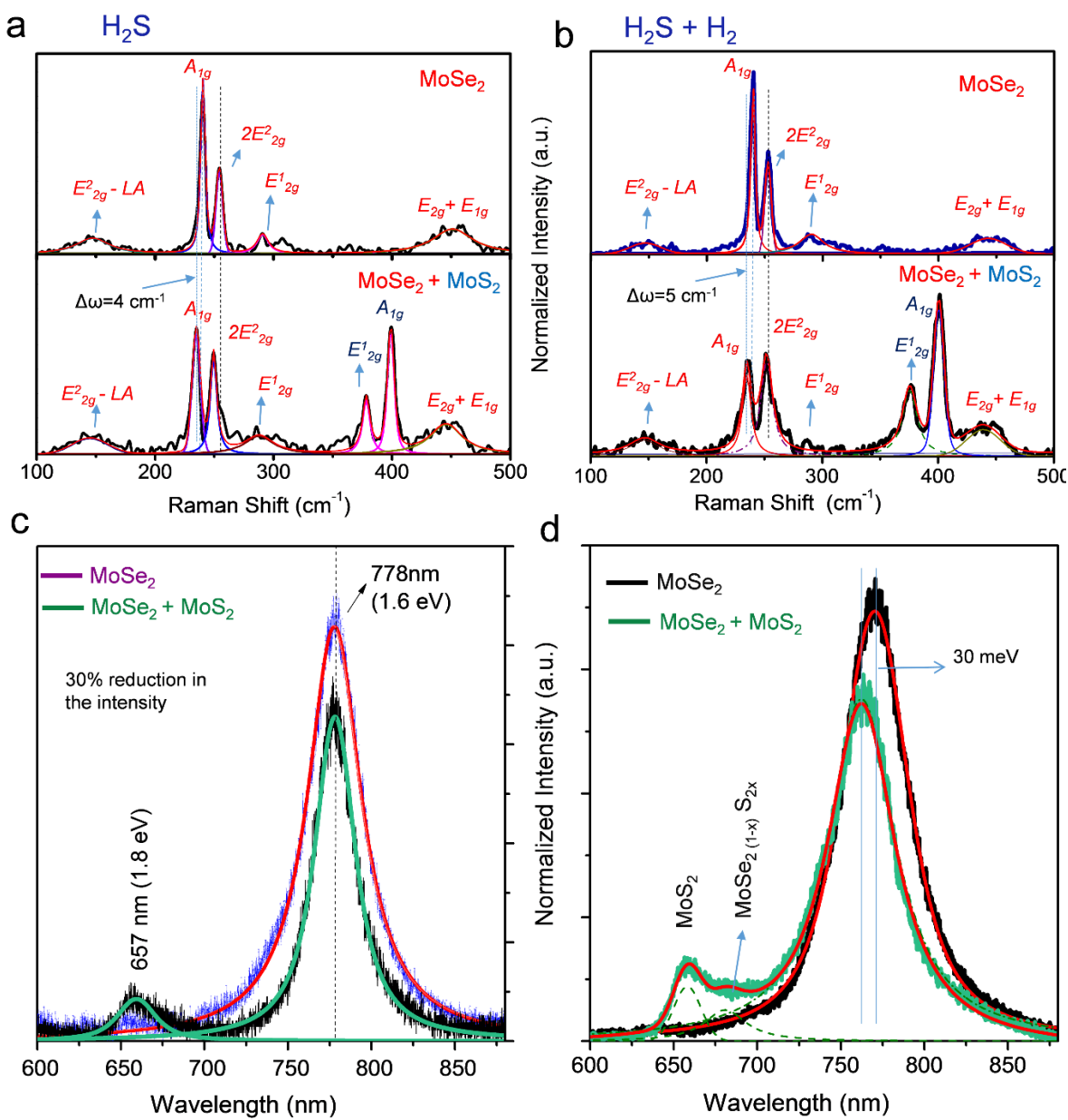

**Figure *S2*** (a) and (b) Raman spectra before (top) and after (bottom) conversion in H$_2$S and H$_2$S+H$_2$ atmospheres, respectively. (c) and (d) PL spectra before and after conversion in H$_2$S and H$_2$S+H$_2$ atmospheres, respectively.

**Table S2**. Parameters obtained from fitting the time-dependent data in Figure S1d, to eq.(1) and eq.(2).

| Gas | Phonon Mode | $\tau_1$ (s) | $\tau_2$ (s) | $\tau_3$ (s) |
|---|---|---|---|---|
| H$_2$S | (MoSe$_2$) A$_{1g}$ | 3 | 333 ± 14 | - |
| | (MoS$_2$) A$_{1g}$ | - | - | 2427 ± 60 |
| H$_2$S+H$_2$ | (MoSe$_2$) A$_{1g}$ | 1.35 | 76 ± 4 | - |
| | (MoS$_2$) A$_{1g}$ | - | - | 846 ± 32 |

***Time dependent chemical reaction and Raman peaks:***

For fixed conditions of laser power, wavelength and TMD film thickness, it is reasonable to assume that the Raman peak intensity of the different phonon modes is directly proportional to the number of related metal-chalcogen (M-X) chemical bonds $[C_X] \propto I_{Raman}$.

In 1$^{st}$ order chemical reaction, the rate of the reaction $(d[C_X]/dt)$ depends linearly on the concentration of the reactant $[C_X]$.

So, for the *selenium release process*:

$$\frac{d[C_{Se}]}{dt} = -k[C_{Se}] \quad \rightarrow \quad \frac{d[C_{Se}]}{[C_{Se}]} = -k\,dt = -\frac{1}{\tau}\,dt$$

The above equation assumes that the composition [$C_{Se}$] decreases asymptotically to zero. However, in our experiments the Raman signal decreases to a finite value different from zero. Suggesting that not all the WSe$_2$ exposed to the laser spot, which contributes to the Raman signal, is converted into a WS$_2$. Due to the Gaussian intensity distribution of the laser, only the material exposed to the highest laser power, within the inner region of the spot, is converted. Hence the final equation has to take this into account. To this end, we define a new relative composition: $[C'_{Se}] = [C_{Se}] - C_{min}$, where $C_{min}$ is the concentration of metal-selenium bonds that were not converted to metal-sulfide, but still contributes to the Raman spectra.

Integrating the concentration between $(C_0 - C_{min})$ and $([C_{Se}] - C_{min})$, where $C_0$ is the initial concentration; and taking the time integration limits between 0 and t:

$$\int_{C_0-C_{min}}^{C_{Se}-C_{min}} \frac{d[C'_{Se}]}{[C'_{Se}]} = -k \int_0^t dt$$

$$\ln[C'_{Se}]_{C_0-C_{min}}^{[C_{Se}]-C_{min}} = -kt$$

$$\ln\left(\frac{[C_{Se}]-C_{min}}{C_0-C_{min}}\right) = -kt$$

$$[C_{Se}] = (C_0 - C_{min})\, e^{-\frac{t}{\tau}} + C_{min}$$

Since the fitting for the experimental data required two exponentials, we might consider that two mechanisms are contributing to the selenium vacancies creation, then there are two time constants involved in the process:

$$\frac{d[C'_{Se}]}{dt} = -k\,[C'_{Se}] = -(k_1 + k_2)\,[C'_{Se}] = -\left(\frac{1}{\tau_1} + \frac{1}{\tau_2}\right)[C'_{Se}]$$

$$k = \frac{1}{\tau} = \frac{1}{\tau_1} + \frac{1}{\tau_2} \quad \text{and} \quad [C_{Se}] = (C_0 - C_{min})\left(e^{-\frac{t}{\tau_1}} + e^{-\frac{t}{\tau_2}}\right) + C_{min}$$

Since $[C_{Se}] \propto I_{WSe_2}$:

$$I_{WSe_2} = (I_0 - I_{min})\left(e^{-\frac{t}{\tau_1}} + e^{-\frac{t}{\tau_2}}\right) + I_{min}$$

### *Sulfur incorporation process*:

The time evolution data for the Raman intensity of WS$_2$ phonon modes was fitted with only one exponential function. From the experiments it can be observed that the relationship between the rate of change of the Raman intensity ($dI_{WS2}/dt$) and the intensity $I_{WS2}$ can be written using a linear dependence:

$$\frac{dI_{WS_2}}{dt} = k\,(I_{max} - I_{WS_2})$$

where, $I_{max}$ is the maximum saturation value for the Raman intensity of these modes. Considering the proportionality $[C_X] \propto I_{Raman}$, we can write the rate equation for the concentration of M-S bonds as:

$$\frac{d[C_s]}{dt} = k\,(C_{max} - [C_s]) = -k\,([C_s] - C_{max})$$

$$\frac{dC_s}{([C_s] - C_{max})} = -k\,dt = -\frac{1}{\tau_3}\,dt$$

$$\int_0^{C_s} \frac{d[C_s']}{([C_s'] - C_{max})} = -\frac{1}{\tau_3}\int_0^t dt$$

$$\ln\,([C_s'] - C_{max})\Big|_0^{C_s} = -\frac{t}{\tau_3}$$

$$\ln\left(\frac{[C_s] - C_{max}}{-C_{max}}\right) = -\frac{t}{\tau_3}$$

$$[C_s] = -C_{max}e^{-\frac{t}{\tau_3}} + C_{max}$$

$$[C_s] = C_{max}\left(1 - e^{-\frac{t}{\tau_3}}\right)$$

Since $[C_s] \propto I_{WS_2}$:

$$I_{WS_2} = I_{max}\left(1 - e^{-\frac{t}{\tau_3}}\right)$$